\documentclass[proceedings]{rmaa}

\usepackage{rmaacite}

\renewcommand{\P}[1]{%
\ifnum#1=1\hbox{OW~168--326E}\fi
\ifnum#1=2\hbox{OW~167--317}\fi
\ifnum#1=3\hbox{OW~163--317}\fi
\ifnum#1=5\hbox{OW~158--323}\fi
\ifnum#1=0\hbox{OW~171--334}\fi}

\title{Getting the Red Out: A Stellar Approach to Determining Spatial 
	Variations of Interstellar Dust}
\author{I. I. Ivans (UT Austin \& McDonald Obs.), C. Sneden (UT Austin 
	\& McDonald Obs.), R. P. Kraft (UCSC \& Lick Obs.), N. B. Suntzeff 
	(CTIO), V. V. Smith (UT El Paso \& McDonald Obs.), G. E. 
	Langer\altaffilmark{1} (Colorado Coll.), and J. P. Fulbright (DAO)}
\altaffiltext{1}{Deceased.}

\fulladdresses{
\item I. I. Ivans and C. Sneden: Department of Astronomy \& McDonald Observatory, University of Texas, Austin, TX 78712 (iivans@astro.as.utexas.edu, chris@verdi.as.utexas.edu)
\item R. P. Kraft: UCO/Lick Observatory, Department of Astronomy \& Astrophysics, University of California, Santa Cruz, CA  95064 (kraft@ucolick.org)
\item N. B. Suntzeff: Cerro Tololo Inter-American Observatory, National Optical Astronomy Observatories, operated by the Association of Universities for Research in Astronomy, Inc., (AURA), under cooperative agreement with the National Science Foundation.  La Serena, Chile (nsuntzeff@noao.edu)
\item V. V. Smith: Department of Physics \& McDonald Observatory, University of Texas at El Paso, 500 West University, El Paso, TX 79968-0515 (verne@balmer.physics.utep.edu)
\item G. E. Langer: Physics Department, Colorado College, Colorado Springs, CO
80903 (Deceased 1999 Feb. 16)
\item J. P. Fulbright: Herzberg Institute of Astrophysics, Dominion Astrophysical Observatory, Victoria, BC, Canada V8X 4M6 (jon.fulbright@hia.nrc.ca)}

\shortauthor{Ivans et al.}
\shorttitle{Getting the Red Out}

\keywords{ISM: dust, extinction --- ISM: structure --- stars: fundamental 
	parameters --- globular clusters: general --- globular clusters: 
	individual (NGC~6121)}

\abstract{By employing two different spectroscopic techniques we have 
mapped out the variable ISM dust extinction endemic to globular cluster
M4.  We derive an average $E(B-V)$ reddening of 0.33 $\pm$ 0.01 and 
$R = 3.4 \pm 0.4$, both in good agreement with previous studies of M4.  
For individual stars in the most heavily reddened regions of M4, we find 
$E(B-V)$ values significantly higher than those inferred by studies of 
interstellar gas---for heavily reddened regions, the gas measured in 
column density measurements may not completely trace the dust.}

\listofauthors{I.~I.~Ivans, C.~Sneden, R.~P.~Kraft, N.~B.~Suntzeff, 
	V.~V.~Smith, G.~E.~Langer \& J.~P.~Fulbright}
\indexauthor{Ivans, I.~I.}
\indexauthor{Sneden, C.}
\indexauthor{Kraft, R.~P.}
\indexauthor{Suntzeff, N.~B.}
\indexauthor{Smith, V.~V.}
\indexauthor{Langer, G.~E.}
\indexauthor{Fulbright, J.~P.}

\begin{document}

\maketitle

\section{Introduction}
\label{sec:intro}
\indent How much light from a particular astronomical source is attenuated 
due to intervening dust?  How does the dust obscuration vary on spatial 
scales?  How well do standard techniques used for studying interstellar 
gas infer the reddening due to dust along a particular line-of-sight?  Using 
observations of individual stars in globular cluster M4, we have developed a 
technique which directly recovers the intrinsic reddening information of the 
intervening dust.

\begin{figure}
  \begin{center}
    \leavevmode
    \includegraphics[height=3.40in]{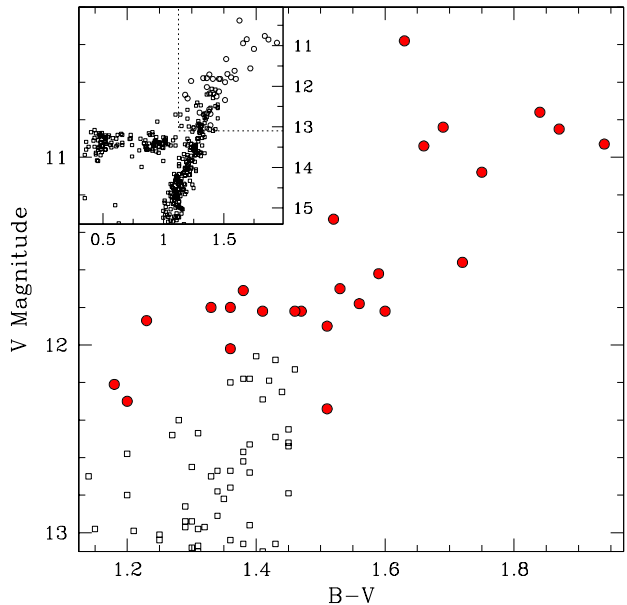}
    \caption{A colour-magnitude diagram of M4, with photometry from Cudworth 
	\& Rees (1990) showing the positions of our program stars in solid 
	circles.   The inset diagram shows the program stars with respect to 
	all of the Cudworth \& Rees M4 stars of magnitude $\le$ 15.5.}
    \label{fig:hr.ps}
  \end{center}
\end{figure}

\indent M4 is essentially the closest globular cluster, located at a distance 
of roughly 2~kpc.  The line of sight to M4 passes through the outer parts of 
the Scorpius-Ophiuchus dark cloud complex, which lies $\sim$150 to 250 pc 
\cite{CR90} from the Sun.  Visual extinction is high, making this region 
exceptionally difficult to study in the optical. The interstellar reddening 
in this region is also differential; the reddening varies across the face of 
M4 (Cudworth \& Rees 1990; Liu \& Janes 1990; Minniti et al. 1992).  
And, the dust extinction probably varies on small spatial scales as well.  
This is suggested by the scatter in the colour-magnitude diagram of M4 (see 
Figure~\ref{fig:hr.ps}): the subgiant and giant branches are much broader 
than can be attributed to photometric uncertainties.  In this paper, we 
describe a stellar approach to deriving $E(B-V)$ and $R$ 
[$\equiv E(V-K)/E(B-V) = A_{V}/E(B-V)$] for this cluster using high
resolution ($\Delta\lambda/\lambda$ = 50-60,000) data taken with 
cross-dispersed spectrographs  at the Lick (Vogt 1987, Valenti et al. 1995)
and McDonald Observatories (Tull et al. 1995).

\section{Stellar Parameters}

\indent We combined traditional spectroscopic abundance methods with
modifications to the line-depth ratio technique pioneered by Gray (1994) to 
determine the atmospheric parameters of our stars.  The ``Gray'' method 
relies on ratios of central depths of carefully selected absorption
features which have different functional dependences on effective 
temperatures (eg. vanadium versus neutral or ionized iron; see Gray 1994
table 2) to derive accurate relative temperature rankings .  Gray's work was 
done on main sequence stars similar to the Sun in metallicity and has since 
been expanded by Hiltgen (1996) for applications to subgiants of a range of 
disk metallicities. Happily, many of Gray's line depth ratios are also 
sensitive Teff indicators for lower metallicity very cool red giant branch 
(RGB) stars.  Within most clusters, there is no evidence that there may be 
any important variations in star-to-star abundances of the elements producing
the temperature-sensitive lines.  The line depth ratios vary more than one 
dex in spectra of giants of moderately metal-poor clusters, and thus can 
indicate very small Teff changes.  

\indent Our initial Teff calibration of the M4 line depth ratios was set 
through a similar analysis of RGB stars of M5.  M5 is a cluster of very 
similar metallicity but suffers little from interstellar dust extinction.  
The details of the correlations and transformations are discussed in Ivans et 
al. (1999; \S 3).  In brief, we derived fits of the measured ratios in M5 
stars to their temperatures and colour indices.  We then inverted the 
relations to predict relative Teffs in our M4 stars using the M4 line depth
ratios.  That is, the ``Gray'' method permitted us to accurately rank our 
stars using M5 as the temperature calibrator.  Then, we relied on a full 
spectral analysis to determine the actual M4 star temperatures.   Subsequent 
iterations of the models yielded final Teffs that were obtained by ensuring 
that no trends of abundances were derived from individual Fe~{\sc i} lines 
as a function of the excitation potentials of the individual lines. 
Combining the two spectroscopic methods permitted us to bypass the 
difficulties inherent in using the M4 $B-V$ photometry to determine the 
model atmosphere parameters.  

\section{Disentangling the Dust}

\indent Taking advantage of the non-photometric means by which we obtained 
our temperatures, we estimated the intrinsic $B-V$ colours by interpolating 
the model atmosphere predictions of Cohen et al. (1978) and 
subtracting these intrinsic $B-V$ colours from the observed colours.  The
results for individual stars can be found in Ivans et al. (1999; \S 3). 
We then made star-by-star comparisons of our results against $E(B-V)$ 
estimates derived independently in interstellar absorption studies of 
potassium by Lyons et al. (1995).  

\indent For three particular stars, our reddening estimates are 
significantly higher than those derived by Lyons et al.  These stars are in 
the western half of the cluster, the most heavily reddened region of M4, 
according to previous differential reddening results (\S \ref{sec:intro}) 
and a region on the sky for which higher than average IRAS 100~\micron\ flux 
values have also been measured. The stars we observed nearest to these in the 
sky also have relatively high reddening values.  It may simply be that the 
gas measured in the K~{\sc i} column density measurements---the basis of the 
Lyons et al. reddening estimates---does not completely trace the dust or the 
IRAS flux measurements, due to shielding effects and variations in the 
optical depth of the gas as suggested by de~Geus \& Burton (1991) in their 
study of the Sco-Oph molecular cloud region.  With the exception of the 
values derived for stars in the region of highest IRAS fluxes, where 
standard interstellar methods seem to underestimate the values, the 
agreement for the remainder of the stars in common is excellent: 
$<\delta E(B-V)>$~= +0.01~$\pm$~0.01~magnitude ($\sigma$~= 0.05~magnitude).

\indent Our average $E(B-V)$ reddening of 0.33 $\pm$ 0.01, while in good 
agreement the M4 RR Lyrae studies by Caputo et al. (1985; $E(B-V)$~= 
0.32--0.33), is significantly lower than that estimated by using the dust 
maps made by Schlegel et al. (1998).  The difference is 
$<$$\delta$E(B--V)$>$~= --0.17~$\pm$~0.01~magnitudes ($\sigma$~= 
0.04~magnitude).  Similar differences have been found for the Taurus
dark cloud complex (Arce \& Goodman 1999) and for galaxy WKK5345 (Woudt 
1998), close to the galactic plane.  Arce \& Goodman attribute the cause of 
the dust opacity estimate discrepancy to the correlation of intrinsic 
$E(B-V)$ and Mg$_{2}$ assumed by Schlegel et al., which produces an overall 
overestimate of reddening values in regions of large extinction.

\indent In addition to large and differential reddening in M4, there is 
evidence that the dust along the line of sight has anomalous absorption 
properties that deviate from the ``normal'' law of interstellar extinction 
described by $R$ values in the range of 3.0 to 3.1 (Harris 1973, Barlow \& 
Cohen 1977, Sneden et al. 1978).  Studying the outer parts of the Sco-Oph 
dust cloud complex, Chini (1981) determined $R = 4.2 \pm 0.2$ in this region 
of the sky.  Cudworth \& Rees (1990) found $R=3.3 \pm 0.7$.  Work done by 
Clayton \& Cardelli (1988) gives $R$~=~3.8 for $\sigma$ Sco, a star only one 
degree in the plane of  the sky away from M4.  Many other investigations, 
with similar, high $R$ results, are reviewed by Dixon \& Longmore (1993).  

\indent To examine this issue, we determined the intrinsic $V-K$ colours by 
interpolating the model atmosphere predictions  of Cohen et al. 
(1978) and comparing the predicted values against the un-dereddened 
observations of Frogel et al. (1983).  Combined with our derived $E(B-V)$
values, we obtained a value of $R = 3.4 \pm 0.4$, in reasonable agreement 
with that of previous M4 studies.

\section{Summary, Discussion, and Future Work}
Confronted with a cluster having large and variable interstellar extinction 
across the cluster face, we combined traditional spectroscopic abundance 
methods with modifications to the line-depth ratio technique pioneered by 
Gray (1994) to determine the atmospheric parameters of our stars.  We derive
 a total-to-selective extinction ratio of 3.4~$\pm$~0.4 and an average 
$<$E(B--V)$>$ reddening of 0.33~$\pm$~0.01, in good agreement with previous 
M4 studies but significantly lower than that estimated by using the dust maps 
made by Schlegel et al. (1998).

\indent We find that the gas measured in the gas column density 
measurements seems to underestimate the dust contribution in regions of 
large reddening. In the case of M4, the stellar approach probes a larger 
dynamic range of reddening values than interstellar absorption line 
techniques yield. However, because our technique relies on the broad-band 
photometry, our approach does not yield information about individual 
reddening components along the line of sight. Thus, for studies of
interstellar dust, the stellar approach cannot supplant the techniques of 
studying interstellar gas, but does complement them.

\indent A comprehensive investigation of the spectroscopic line ratios of 
other globular clusters, utilizing high resolution data, is underway and 
results will be reported in a future publication. 

\acknowledgements We are indebted to Jerry Lodriguss for generously sharing 
the electronic files containing his excellent deep sky images of the M4 
Sco-Oph region.  David Schlegel has our appreciation for answering questions 
regarding the dust map code as does David Gray for providing helpful 
feedback regarding the applicability of the spectroscopic line ratio 
technique.  We are grateful to Renee James for doing some of the initial
work in getting the line ratio technique ready to apply to the cluster stars
and to  Flavio Fusi Pecci \& Francesco Ferraro for sharing their photometry 
of M4.  IRAS fluxes were obtained using IBIS, an observational planning tool 
for the infrared sky developed at IPAC.  The Infrared Processing and 
Analysis Center is operated by the California Institute of Technology and 
the Jet Propulsion Laboraory under contract to NASA.  This project has also 
made use of NASA's Astrophysics Data System Abstract Service. This research 
was supported by NSF grants AST-9618351 to RPK, AST-9618364 to CS, and 
AST-9618459 to VVS. 

Ed Langer, our co-author, passed away on Feb. 16, 1999, following a brief 
illness.  Ed's kindness and generous spirit added a great deal to our 
collaboration; his way of thinking added a great deal to the science.  He is 
greatly missed.

\end{document}